\documentclass[conference]{IEEEtran}
\IEEEoverridecommandlockouts
\usepackage{cite}
\usepackage{amsmath,amssymb,amsfonts}
\usepackage{algorithmic}
\usepackage{graphicx}
\usepackage{tabularx}
\usepackage{textcomp}
\usepackage{xcolor}
\usepackage{stfloats}
\usepackage{multirow}
\def\BibTeX{{\rm B\kern-.05em{\sc i\kern-.025em b}\kern-.08em
    T\kern-.1667em\lower.7ex\hbox{E}\kern-.125emX}}
\begin{document}

\title{A Novel 3D Space-Time-Frequency Non-Stationary Channel Model for 6G THz Indoor Communication Systems}

\author{Jun Wang\textsuperscript{1,3}, Cheng-Xiang Wang\textsuperscript{1,3,*}, Jie Huang\textsuperscript{1,3}, Haiming Wang\textsuperscript{2,3}
\\
\textsuperscript{1}National Mobile Communications Research Laboratory, School of Information of Science and Engineering, 
\\Southeast University, Nanjing 210096, China.\\

\textsuperscript{3}State Key Laboratory of Millimeter Waves, School of Information of Science and Engineering, 
\\Southeast University, Nanjing 210096, China.\\
\textsuperscript{2}Purple Mountain Laboratories, Nanjing 211111, China.\\
\textsuperscript{*}Corresponding Author: Cheng-Xiang Wang
\\
Email: \{jun.wang, chxwang, j\_huang, hmwang\}@seu.edu.cn
}

\maketitle

\begin{abstract}
Terahertz (THz) communication is now being considered as one of possible technologies for the sixth generation (6G) communication systems. In this paper, a novel three-dimensional (3D) space-time-frequency non-stationary massive multiple-input multiple-output (MIMO) channel model for 6G THz indoor communication systems is proposed. In this geometry-based stochastic model (GBSM), the initialization and evolution of parameters in time, space, and frequency domains are developed to generate the complete channel transfer function (CTF). Based on the proposed model, the correlation functions including time auto-correlation function (ACF), spatial cross-correlation function (CCF), and frequency correlation function (FCF) are investigated. The results show that the statistical properties of the simulation model match well with those of the theoretical model. The stationary intervals at different frequencies are simulated. The non-stationarity in time, space, and frequency domains is verified by theoretical derivations and simulations.
\end{abstract}


\section{Introduction}
With the rapid development of wireless communication, the data traffic is expected to grown exponentially in the 6G communication system. THz communication is considered as one of the most important key technologies for 6G communication systems. THz wave from 0.1 THz to 10 THz have the ability to provide large bandwidth of more than one hundred gigahertz(GHz)\cite{RN532}. As a result, THz communication can theoretically achieve ultra-high transmission rate of 100 Gbps or even higher\cite{RN300}. %THz communication can be applied to many scenarios that require high data rate such as data center and device-to-device (D2D) communication. Indoor communication is a very important scenario whose requirement of data rate will explosively increase in the future. THz communication can fit the requirement perfectly. 

For the design and optimization of THz communication systems, a THz channel model that can accurately reflect THz characteristics is necessary. An accurate channel model is also the prerequisite for performance evaluation of the communication systems such as capacity analysis\cite{Ge1} and energy efficiency evaluation\cite{Ge3}. Traditional channel models for lower frequency can not be applied to THz band due to some unique characteristics in THz band such as path loss and propagation properties. Path loss and atmospheric absorption of THz channels were investigated in \cite{RN309,RN351,RN256}. The gas absorption caused by oxygen and water vapor will rapidly increase when the frequency becomes higher. %The precise absorption model was given in~\cite{RN309}. 

Propagation properties of THz waves were studied in the literature. 
%It is well known that non-line-of-sight (NLOS) transmissions can be categorized into specularly reflected propagation, diffusely scattered propagation, and diffracted propagation. All the three categories are investigated in THz band.
 In \cite{RN308,RN356}, measurement and modeling of multiple reflection effects in building materials at THz were introduced. The reflection loss shows great dependence on frequency and materials and can be calculated by Kirchhoff theory. According to the measurement in \cite{RN197}, high-order paths are very hard to be detected due to high reflection loss. The diffusely scattered propagation was investigated in\cite{RN364,RN294,RN346,RN369}. %The wavelength of THz wave is 
 %less than 1 mm when the frequency is higher than 300 GHz. It is
 % not much larger than the roughness of some common materials. As a result, 
 In THz band, more power is diffusely scattered when frequency increases. %and the specular reflection power is not dominated. 
 All the diffusely scattered rays happen around the specular reflected path and scattering remote from the immediate region around the specular reflection points can be neglected during the channel modeling. %In \cite{RN310} and \cite{RN349}, diffraction of THz wave was studied with different materials. It can be concluded that diffraction at edges does not play a significant role as they are almost everywhere in a room. 
  However, most of the measurements of channel characteristics were carried out at 300 GHz, channel characteristics of reflection and scattering in higher frequencies need to be investigated experimentally. 

Many THz channel models were proposed for indoor THz communications. In \cite{RN207}, a novel channel model based on ray tracing was proposed, which incorporated the propagation models for the line-of-sight (LOS), reflected, scattered, and diffracted paths. This channel model was validated with the experimental measurements from the literature. In \cite{RN194}, a stochastic channel model for indoor scenarios considering frequency dispersion was proposed. However, channel models based on ray tracing methods are not general and flexible. The existing stochastic THz channel models cannot show the unique propagation characteristics of THz waves in the indoor scenario and they do not support dynamic environments. In addition, some non-stationary channel models for millimeter wave band communication systems were proposed in \cite{RN321,RN512,RN322,RN161}. However, the evolution of clusters in these channel models are not suitable for THz band channel models. 

In this paper, a 3D space-time-frequency non-stationary GBSM for THz communication systems is proposed. The mobility of users and ultra massive MIMO are considered for THz indoor communication systems to model the non-stationarity in time and space domains. In frequency domain, we consider that diffusely scattering in a surface is frequency dependent according to the investigation of THz propagation.
%In this THz channel model, %we consider both the line-of-sight (LOS) and the NLOS path which including single bounced paths and multi bounced paths. % Each cluster respresent a specular reflection point which is surrounded by scattered points.  
 %It is assumed that the Tx and the Rx are equipped with directional uniformly linear antenna arrays. % and the speed of Tx or Rx is .  
To obtain the space-time-frequency dependent channel coefficient, we will first initialize the channel at first element of transmitter (Tx) and receiver (Rx), time $t_0$, and frequency $f_0$. Then the channel evolution will be taken in time, space, and frequency domains. After the space-time-frequency dependent CTF is generated, channel characteristics such as delay power spectrum density (PSD), ACF , CCF and FCF are studied. 

The remainder of this paper is organized as follows. Section~\uppercase\expandafter{\romannumeral2} describes the proposed GBSM in details and gives the CTF of the channel. In Section~\uppercase\expandafter{\romannumeral3}, channel characteristics are calculated. Then simulation results are compared and analyzed in Section~\uppercase\expandafter{\romannumeral4}. Finally, conclusions are drawn in Section~\uppercase\expandafter{\romannumeral5}.

\section{A Novel 3D THz MIMO GBSM}

\subsection{Description of the Channel Model}

Let us consider a MIMO indoor communication system equipped with $M_T$ transmit and $M_R$ receive antennas. The center frequency is $f_c$. Let $\text{Ant}_p^T$ denote the $p$th transmitted antenna and $\text{Ant}_q^R$ denote the $q$th received antenna. It should be noted that arbitrary antenna array layouts can be considered in the proposed model.
%Typical antenna array layouts include uniform linear arrays, 2D planar arrays, and 3D cube arrays. Antenna responses can be modified subject to actual antenna settings. 

%In THz communication, the unique propagation characteristics need to be considered. Scattering becomes more and more important as the frequency increase. In this model, we assume that the each specular reflection point is surrounded by several scattering points to imitate the diffusely scattering. The deviation of these scattering points is decided by the frequency and surface materials. Scattering from one surface is considered as a cluster. For a certain surface, the roughness is fixed. Therefore, the delay dispersion and angle dispersion in each cluster are only related to frequency.
%due to high resolution in THz band. however, scattering at one surface may cause angle dispersion and can be considered as a cluster. The scattered points are distributed arround the speclar reflection point. The range of scattered points is related to the frequency and the metarial of the reflected surface so the relative angle intra a cluster is changing via frequency.

The proposed non-stationary THz GBSM is illustrated in Fig. 1. In THz band, the wavelength of the carrier frequency is less than one millimeter and comparable to the roughness of some common materials. In this model, each cluster is comprised of diffuse sacttering rays from one roughness surfaces. The center of cluster is considered  as specular reflected point. Notice that coordinate ($x_G$, $y_G$, $z_G$) is established as the global coordinate system (GCS) whose origin is at the first element of the transmit antenna array. This needs to be distinguished from the local coordinate system (LCS) whose origins are at the centers of transmit and receive antenna arrays when calculating 3D radiation pattern. %In this model, omni directional antenna arrays are assumed. 

\begin{figure}[tb]
	\centerline{\includegraphics[width=0.5\textwidth]{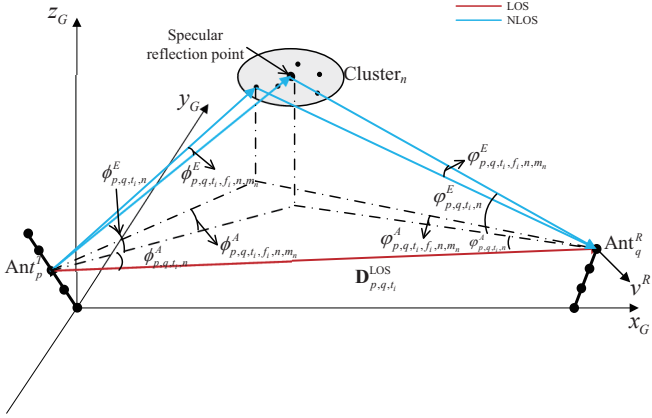}}
	\caption{A 3D THz GBSM for indoor communication.}
	\label{fig_1}
\end{figure}

%All the paths are composed of LOS(line of sight) and NLOS(non line of sight). The NLOS part is only considered to only contain reflection including specluar reflection and diffuse scattering. The diffraction in THz band is so week compared to reflction that can be neglected in the channel model. For Each path of nonLOS, it contains first bounce cluster and last bounce cluster. It should be noted that for single bounce path, the first bounce cluster and last bounce cluster are the same cluster. 

Let $\textbf{A}_{p,t_i}^T$ and $\textbf{A}_{q,t_i}^R$ denote the position vectors of $\text{Ant}_p^T$ and $\text{Ant}_q^R$ at time $t_i$, respectively. Also, let $\psi_A^T$ and $\psi_E^T$ be azimuth and elevation angles of the transmit array, and let $\psi_A^R$ and $\psi_E^R$ be azimuth and elevation angles of the receive array, respectively. $\textbf{D}$ represents the initial position vector of the receiver and it is assumed to be $[D_0, 0, 0]^T$. \textit{$D_0$} is the initial distance between the first elements of Tx and Rx at time $\textit{t}_{0}$. 

The LOS distance vector $\textbf{D}_{p,q,t_i}^{\text{LOS}}$ is computed as 
\begin{equation}\label{D_LOS}
\textbf{D}_{p,q,t_i}^{\text{LOS}}=\textbf{D}+\textbf{A}_{q,t_i}^R-\textbf{A}_{p,t_i}^T+\textbf{v}^R(t_n-t_0).
\end{equation}
It is clear that $\textbf{D}$ equals $\textbf{D}_{1,1,t_0}^{\text{LOS}}$. Also, the distance between $\text{Ant}_p^T$ and $\text{Ant}_q^R$ at time $t_i$ is ${D}_{p,q,t_i}^{\text{LOS}}=\left \|\textbf{D}_{p,q,t_i}^{\text{LOS}}\right \|$, where $\left \|\cdot\right \|$ calculates the Frobenius norm.
% Here, the receive antenna array is assumed to move so that the Dopplar frequency is still considered. 
 %For SB paths and MB paths, the clusters and scatterers are independent. The positions of the clusters are generated according to the category. The first kind of clusters are uniformed distributed in two spheres, denoted as S1 and M1 for SB and MB respectively. The second kind of clustered by ray tracing. We only calculated six specular reflection points for six surfaces of the room with the position of Tx and Rx. For multi bounce path, these cluster in the surfaces of the room are generated randomly. For the last kind, clusters are generated by homogeneous Poisson point process for both SB and MB paths. After we generate the clusters, we use Poisson cluster process to generate scatterers for each cluster around them. 
 
 %\begin{spacing}{2}
 \renewcommand{\arraystretch}{1.3}
 \begin{table*}[tb]
 	\centering
 	\caption{Definition of Main Parameters.}
 		\begin{tabular}{|l|l|}			

			\hline
 			\textbf{Parameters}&\textbf{Definition}\\
 			\hline
 			\textbf{D}& Initial 3D distance between the $T_x$and the $R_x$ at $t_0$\\
 			\hline
 		%	$D_0$& Initial distance between the $T_x$and the $R_x$ at $t_0$
 		%	\\\hline
 			$\textbf{A}_{p,t_i}^T$, $\textbf{A}_{q,t_i}^R$& 3D vector from the first element of transmit array/receive array at $t_0$ to $\text{Ant}_p^T$, $\text{Ant}_q^R$ at $t_i$
 			\\\hline
 			$\textbf{v}^T$, $\textbf{v}^R$&3D velocity vector of transmit and receive array
 			\\\hline
 			$\textbf{D}_{p,q,t_i}^{LOS}$&3D distance vector of the LOS component between $\text{Ant}_p^T$ and $\text{Ant}_q^R$ at $t_i$
 			\\\hline
 			${D}_{p,q,t_i}^{LOS}$&Distance vector of the LOS component between $\text{Ant}_p^T$ and $\text{Ant}_q^R$ at $t_i$
 			\\\hline
 			$\phi_{p,q,t_i}^{A,LOS}$, $\phi_{p,q,t_i}^{E,LOS}$&Azimuth and elevation angles of departure between $\text{Ant}_p^T$ and $\text{Ant}_q^R$ at $t_i$
 			\\\hline
 			$\varphi_{p,q,t_i}^{A,LOS}$, $\varphi_{p,q,t_i}^{E,LOS}$&Azimuth and elevation angles of arrival between $\text{Ant}_p^T$ and $\text{Ant}_q^R$ at $t_i$
 			\\\hline
 			$N$&Total number of observable NLOS clusters
 			\\\hline
 			$M_n$&Numbers of rays within $\text{Cluster}_n$
 			\\\hline
 			$\tilde{A}_{p,q,t_i,n}$&Mirror position of $\text{Ant}_{p,t_i}^T$ for $\text{Cluster}_n$ at $t_i$
 			\\\hline
 			$\tilde{\textbf{D}}_{p,q,t_i,n}$& 3D vector from $\tilde{A}_{p,q,t_i,n}$ to $\text{Ant}_q^R$ for $\text{Cluster}_n$ at $t_i$
 			\\\hline
 		%	$D_{p,q,t_i,n}$& distance between $\text{Ant}_p^T$ and $\text{Ant}_q^R$ via $\text{Cluster}_n$ at time $t_i$ 
 		%	\\\hline
 			$P_{p,q,t_i,f_i,n}$&Mean power of the $n$th cluster between $\text{Ant}_p^T$ and $\text{Ant}_q^R$  at $t_i$ and $f_i$
 			\\\hline
 			$\tau_{p,q,t_i,n}$&Delay from $\text{Ant}_p^T$ to $\text{Ant}_q^R$ via $\text{Cluster}_n$ at $t_i$
 			\\\hline
 			$\tau_{p,q,t_i,f_i,n,m_n}$&Relative delay from $\text{Ant}_p^T$ to $\text{Ant}_q^R$ via $m$th ray in $\text{Cluster}_n$ at $t_i$ and $f_i$
 			\\\hline
 			$\phi(\varphi)_{p,q,t_i,n}^A,\phi(\varphi)_{p,q,t_i,n}^E$&Azimuth and elevation angles of departure(arrival) between $\text{Cluster}_n$ and $\text{Ant}_p^T$ at $t_i$
 			\\\hline
 		%	$\varphi_{p,q,t_i,n}^A,\varphi_{p,q,t_i,n}^E$&azimuth and elevation angles of arrival between $\text{Cluster}_n$ and $\text{Ant}_q^R$ at $t_i$
		%	\\\hline
			$\phi(\varphi)_{p,q,t_i,f_i,n,m_n}^A,\phi(\varphi)_{p,q,t_i,f_i,n,m_n}^E$&Azimuth and elevation angles of departure(arrival) between $m_n$th ray of $\text{Cluster}_n$ and $\text{Ant}_p^T$ at $t_i$ and $f_i$
		%	\\\hline
		%	$\varphi_{p,q,t_i,f_i,n,m_n}^A,\varphi_{p,q,t_i,f_i,n,m_n}^E$&azimuth and elevation angles of arrival between $m_n$th ray of $\text{Cluster}_n$ and $\text{Ant}_q^R$ at $t_i$ and $f_i$

			\\\hline
 	
 		\end{tabular}

 		\label{tab1}
 \end{table*}
%	\end{spacing}
\subsection{The Theoretical Model}

In this channel model, the LOS and non-LOS (NLOS) components are considered. The NLOS components are combined of single bounce and double bounce paths. Higher order paths are neglected due to very high reflection loss.

%Because of the reflection loss in THz band is very high, we only consider two, three and four order reflection paths. Higher order reflection is neglected in this model. For each multi reflection links, we will generate two clusters as the first and last bounce of the paths. The links between the first and last bounce are considered as virtual link.

Considering the space-time-frequency non-stationarity, the CTF at time $t_i$ and frequency $f_i$ can be characterized by an $M_T\times M_R$ matrix $\textbf{H}_{t_i,f_i}(f)=[H_{p,q,t_i,f_i}(f)]_{M_T\times M_R}$. The element of the matrix means the CTF for $p$th transmit element and $q$th receive element, and can be expressed as 

%The channel model consists of two components, i.e., the LOS component and the non-LOS (NLOS) component, and can be written as 

\begin{equation}\label{eq_CTF}
\begin{split}
&\hspace{3mm}H_{p,q,t_i,f_i}(f)=H_{p,q,t_i}^{\text{LOS}}(f)\\&+\sum_{n=1}^{N}\sum_{m_n=1}^{M_n}\lim_{M_n\rightarrow\infty}H_{p,q,t_i,f_i,n,m_n}^{\text{NLOS}}(f)
\end{split}
\end{equation}
where \textit{N} is number of clusters including single bounce and double bounce clusters, $M_n$ is the number of rays within $\text{Cluster}_n$.
% $\tau_{p,q,t_i,f_i,n}$ is the delay of $cluster_n$, and $\tau_{p,q,t_i,f_i,n,m_n}$ is the relative delay of the $m_n$th ray in $cluster_n$. 
In this model, \textit{N} is generated randomly, but it is a constant during
the generation of channel coefficients. The subscript $t_i$ and $f_i$ mean specific time and frequency. 
%The NLOS part is divided to single bounce and multi bounce paths. The \textit{N} of is generrated for single bounce and multi bounce paths. Also, 

\subsubsection{LOS}
For the LOS component, if polarized antenna arrays are assumed at both the Tx and Rx sides, the complex channel coefficient $H_{p,q,t_i}^{\text{LOS}}(f)$ is presented as
% (\ref{h_LOS}) at the bottom of this page 
\begin{equation}
\label{h_LOS}
\begin{split}
&H_{p,q,t_i}^{\text{LOS}}(f)=\\
&\left[ 
\begin{matrix}
F_{p,V}^T(\phi_{E,p,q,t_i}^{\text{LOS}},\phi_{A,p,q,t_i}^{\text{LOS}},f_i)\\
F_{p,H}^T(\phi_{E,p,q,t_i}^{\text{LOS}},\phi_{A,p,q,t_i}^{\text{LOS}},f_i)
\end{matrix}\right]^T 
\left[
\begin{matrix}
e^{j\Phi_{\text{LOS}}}&0\\
0&-e^{j\Phi_{\text{LOS}}}
\end{matrix}
\right]\\
&\left[ 
\begin{matrix}
F_{p,V}^T(\varphi_{E,p,q,t_i}^{\text{LOS}},\varphi_{A,p,q,t_i}^{\text{LOS}},f_i)\\
F_{p,H}^T(\varphi_{E,p,q,t_i}^{\text{LOS}},\varphi_{A,p,q,t_i}^{\text{LOS}},f_i)
\end{matrix}\right]
\sqrt{P_{p,q,t_i,f_i}^{\text{LOS}}}e^{-j2\pi f\tau_{p,q,t_i}^{\text{LOS}}}
\end{split}
\end{equation}
 where $\Phi_{\text{LOS}}$ is uniformly distributed within (0, 2$\pi$). The superscripts V and H denote vertical polarization and horizontal polarization, respectively. 
 Functions $F^T(a,b,f)$ and $F^R(a,b,f)$ are frequency dependent antenna patterns with input angles \textit{a} and \textit{b} in the GCS. The input angles \textit{a} and \textit{b} need
 to be transformed into the LCS to obtain the antenna patterns. 
 $\tau_{p,q,t_i}^{\text{LOS}}$ is the delay from $\text{Ant}_{p,t_i}^T$ to $\text{Ant}_{q,t_i}^R$ at time $t_i$ which is decided by the distance 
 $\tau_{p,q,t_i}^{\text{LOS}}=D_{p,q,t_i}^{\text{LOS}}/c$.  

To make the proposed channel model more realistic, the path loss composes of free space path loss and the atmosphere absorption. Here, the atmosphere absorption is a function of distance, wavelength, and some environment parameters such as temperature and humidity. To simplify the channel model, we assume the atmosphere absorption is only decided by distance and wavelength. The path loss in dB is given by 
\begin{equation}
P_{f_i}^{\text{LOS}}(D)[\text{dB}]=20log_{10}(\frac{4\pi D}{\lambda_i})+L_{a}(D,\lambda_i)+X(\sigma).
\end{equation}
Here, $\lambda_i$ is the wavelength of frequency $f_i$, $20log_{10}(\frac{4\pi D}{\lambda_i})$ is free space path loss, $L_{a}(D,\lambda_i)$ is the atmosphere absorption, and $X(\sigma)$ is the random loss caused by the system. In this model, $P_{p,q,t_i,f_i}^{LOS}$ can be expressed as $P_{p,q,t_i,f_i}^{\text{LOS}}=P_{f_i}^{\text{LOS}}(D_{p,q,t_i}^{\text{LOS}})$.
%The phase of LOS path $\varphi_{\text{LOS}}$ is assumed to be uniformed distribution, i.e. $\varphi_{\text{LOS}}\sim U[0,2\pi]$.
The angle of departure (AoD) is equal to angle of arrival (AoA) for the LOS path. The azimuth/elevation angle of departure (AAoD/EAoD) $\phi_{A,p,q,t_i}^{\text{LOS}}/\phi_{E,p,q,t_i}^{\text{LOS}}$ and azimuth/elevation angle of arrival (AAoA/EAoA) $\varphi_{A,p,q,t_i}^{\text{LOS}}/\varphi_{E,p,q,t_i}^{\text{LOS}}$ can be calculated by the vector $\textbf{D}_{p,q,t_i}^{\text{LOS}}$.

\subsubsection{NLOS}
%The NLOS part contains paths with different reflection orders. The number of each reflection order is limited. In this channel model, only one order and two order paths are considered because of very high reflection loss in THz band. 
The CTF of the $m_n$th ray in $\text{cluster}_n$ is 
\begin{equation}
\begin{split}
\label{h_NLOS}
&H_{p,q,t_i,f_i,n,m_n}(f)=\\&
\left[ 
\begin{matrix}
F_{p,V}^T(\tilde{\phi}_{p,q,t_i,f_i,n,m_n}^E,\tilde{\phi}_{p,q,t_i,f_i,n,m_n}^A,f_i)\\
F_{p,H}^T(\tilde{\phi}_{p,q,t_i,f_i,n,m_n}^E,\tilde{\phi}_{p,q,t_i,f_i,n,m_n}^A,f_i)
\end{matrix}\right]^T \\
&\left[
\begin{matrix}
\sqrt{k_{n,m_n}^{-1}}e^{j\Phi_{n,m_n}^{VV}}&e^{j\Phi_{n,m_n}^{VH}}\\
e^{j\Phi_{n,m_n}^{HV}}&\sqrt{k_{n,m_n}^{-1}}e^{j\Phi_{n,m_n}^{HH}}
\end{matrix}
\right]\\
&\left[ 
\begin{matrix}
F_{p,V}^T(\tilde{\varphi}_{p,q,t_i,f_i,n,m_n}^E,\tilde{\varphi}_{p,q,t_i,f_i,n,m_n}^A,f_i,f_i)\\
F_{p,H}^T(\tilde{\varphi}_{p,q,t_i,f_i,n,m_n}^E,\tilde{\varphi}_{p,q,t_i,f_i,n,m_n}^A,f_i,f_i)
\end{matrix}\right]
\\&
\sqrt{\frac{P_{p,q,t_i,f_i,n}}{M_n}}e^{-j2\pi f(\tau_{p,q,t_i,n}+\tau_{p,q,t_i,f_i,n,m_n})}.
\end{split}
\end{equation}
In (\ref{h_NLOS}), $P_{p,q,t_i,f_i,n}$ is the power of $\text{Cluster}_n$, $\tau_{p,q,t_i,n}$ and $\tau_{p,q,t_i,f_i,n,m_n}$ are delay of $\text{cluster}_n$ and relative delay of $m_n$th ray in $\text{Cluster}_n$, respectively, $k_{n,m_n}$ denotes the cross polarization power ratio, $\Phi_{n,m_n}^{VV}$, $\Phi_{n,m_n}^{VH}$,$\Phi_{n,m_n}^{HV}$, and $\Phi_{n,m_n}^{HH}$ are the initial random phases of the $m_n$th ray in $Cluster_n$ in four polarization directions. 
Parameters $\tilde{\phi}_{p,q,t_i,f_i,n,m_n}^E$, $\tilde{\phi}_{p,q,t_i,f_i,n,m_n}^A$, $\tilde{\varphi}_{p,q,t_i,f_i,n,m_n}^E$, and $\tilde{\varphi}_{p,q,t_i,f_i,n,m_n}^A$ are EAoD, AAoD, EAoA, and AAoA, respectively. It should be noted that $\tilde{\phi}$ means the summation of angle of a cluster and the relative angle. For example, 
\begin{equation}
\tilde{\phi}_{p,q,t_i,f_i,n,m_n}^E=\phi_{p,q,t_i,n}^E+\phi_{p,q,t_i,f_i,n,m_n}^E.
\end{equation}
Similarly, $\tilde{\phi}_{p,q,t_i,f_i,n,m_n}^A$, $\tilde{\varphi}_{p,q,t_i,f_i,n,m_n}^E$, and $\tilde{\varphi}_{p,q,t_i,f_i,n,m_n}^A$ can be calculated similarly.  
The definitions of main parameters are given in Table \ref{tab1}.

\subsection{The Simulation Model}
In the simulation model, the number of rays within a cluster is assumed as infinite in this theoretical model while it is finite in the simulation model which can be expressed as
\begin{equation}\label{eq_CTF}
\begin{split}
&\hspace{3mm}H_{p,q,t_i,f_i}(f)=H_{p,q,t_i}^{\text{LOS}}(f)+\sum_{n=1}^{N}\sum_{m_n=1}^{M_n}H_{p,q,t_i,f_i,n,m_n}^{\text{NLOS}}(f).
\end{split}
\end{equation} 

In the simulation model, the method of equal area (MEA)\cite{Mobile} is used to obtain the discrete AAoDs, EAoDs, AAoAs, and EAoAs.

\subsection{Channel Initialization}
In this subsection, channel initialization is discussed.  All the parameters for NLOS components are generated for \textit{p}=1, \textit{q}=1 at time $t_0$ and frequency $f_0$. After these parameters are generated, evolution of them will be taken which is introduced in the next subsection.

The number of paths is generated randomly for first bounce and second bounce paths, respectively. For first order path, the number $N_{1\text{st}}$ has the probabilities of $p(4)=0.35$ and $p(5)=0.65$. For second order reflection, the number $N_{2\text{nd}}$ ranges from 7 to 13. The total number of paths is constant during the evolution. It is clear that $N=N_{1\text{st}}+N_{2\text{nd}}$.

The initial cluster delay $\tau_{1,1,t_0,n}$ is generated by random variables $\Delta \tau_{i,1\text{st}}$ and $\Delta \tau_{i,2\text{nd}}$\cite{RN194}, where $\Delta \tau_{i,1\text{st}}$ and $\Delta \tau_{i,2\text{nd}}$ is defined as the time interval of arrival between two adjacent clusters for first order cluster and second order cluster, respectively. 
For first cluster, $\Delta \tau_1$ is the time interval compared to the LOS path. So, we have
\begin{equation}
\tau_{1,1,t_0,i} = \begin{cases}
\tau_{1,1,t_0}^{\text{LOS}}+\Delta \tau_{i,1\text{st}},&i=1;\\
\tau_{1,1,t_0,i-1}+\Delta \tau_{i,1\text{st}},&2\leq i\leq N_{1\text{st}}.
\end{cases}
\end{equation}
 
A similar method is used to generate second bounce clusters which can be expressed as 
\begin{equation}
\tau_{1,1,t_0,i} = \begin{cases}
\tau_{1,1,t_0}^{\text{LOS}}+\Delta \tau_{i,2\text{nd}},&i=N_{1\text{st}}+1;\\
\tau_{1,1,t_0,i-1}+\Delta \tau_{i,2\text{nd}},&N_{1\text{st}}+2\leq i\leq N
\end{cases}
\end{equation}
where $\Delta \tau_{i,1st}$ and $\Delta \tau_{i,2nd}$ are negative exponential (NEXP) random variables with parameters $\mu_{\Delta \tau_{i,1st}}$ and $\mu_{\Delta \tau_{i,2nd}}$. The relative delay $\tau_{1,1,t_0,f_0,n,m_n}$ is also considered as a NEXP random variable with the parameter $\mu_{f_i,n}$. 

The power of the cluster is generated according to the distance \cite{RN194}. It can be expressed as 
\begin{equation}
P_{1,1,t_0,i}(\text{dB}) = P_{1,1,t_0}^{\text{LOS}}-\Delta P_{1,1,t_0}^{\text{LOS}}-n_\tau\cdot(\tau_{1,1,t_0,i}-\tau_{1,1,t_0}^{\text{LOS}})+\Delta a_i
\end{equation} 
where $P_{1,1,t_0}^{\text{LOS}}$ is the function value at the LOS delay with respect to the LOS amplitude, and $n_\tau$ is the temporal decay coefficient. Moreover, each
ray deviates from the straight line by the random variable $\Delta a_i$. 

The initial phase of each cluster is considered as uniformly distributed in $\left[-\pi,\pi \right] $. The EAoD, AAoD, EAoA, and AAoA of the initial position and time are considered as independent Gaussian distributions. The relative angles 
$\phi_{1,1,t_0,f_0,n,m_n}^E$, $\phi_{1,1,t_0,f_0,n,m_n}^A$, $\varphi_{1,1,t_0,f_0,n,m_n}^E$, and $\varphi_{1,1,t_0,f_0,n,m_n}^A$ are Gaussian distributed with zero mean and independent variance $\sigma_{f_i,n}$.
 
%After the center of all clusters are determined, we need to consider the intra cluster parameter caused by diffuse scattering. Due to the diffuse scattering is correlated with wavelength and the surface and the surface is seen unchanged, all these intraclusters are dependent to frequency. The relative delay is exponentially distributed random variable where the average delay will increase when the frequency rises. The relative angles are Gussian distributed with the frequency dependent simga.   
\begin{figure}[tb]
	\centerline{\includegraphics[width=0.4\textwidth]{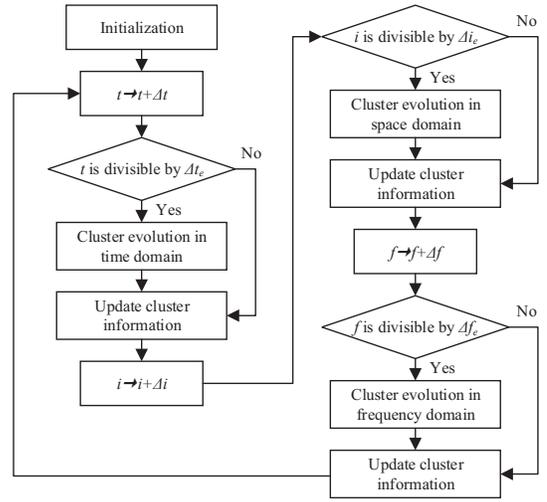}}
	\caption{The flowchart of space-time-frequency cluster evolution of the
		proposed THz channel model.}
	\label{fig_2}
\end{figure}
\subsection{Evolution of Clusters}
In this subsection, the space, time, and frequency domain cluster evolutions for the proposed THz channel model are demonstrated. The flow chart is shown in Fig. \ref{fig_2}.  %In some existed channel models, the evolution of clusters is based on birth-death process. However, birth-death process is not suitable for THz band.
 %because the scatterers is too small compared to lower frequency and the velocity in indoor scenario is quite small in time domain. 
Here, we update the parameters for each cluster in time and space domains based on the geometric relationship. Evolution in frequency domain is also considered as the regeneration of relative distance and relative angle in different frequency points. 

\subsubsection{Evolution in the Time Domain}
\begin{figure}[tb]
	\centerline{\includegraphics[width=0.5\textwidth]{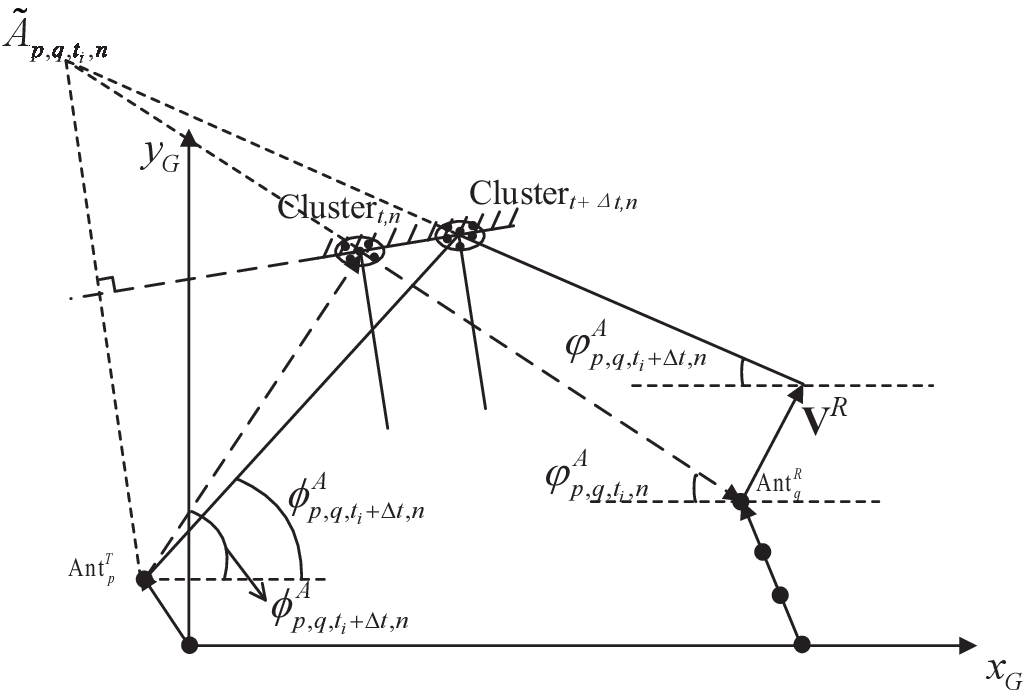}}
	\caption{Illustration of evolution in the time domain.}
	\label{fig_3}
\end{figure}

The Rx is assumed to move to imitate human activity in an indoor office. Usually, the velocity of Rx is small. In a small time interval, the difference of positions is small. We assume that the specular reflection point is still on the same surface, but the position is moving a small distance. Here, we use a single reflection as an example and the geometric relationship is shown in Fig. \ref{fig_3}.

Let us use $\tilde{A}_p^T$ to denote the mirror point of Tx point reflection and $\textbf{D}_{t+\Delta t,p,q,f,n}$ represent the virtual vector from $\tilde{A}_p^T$ to $\text{Ant}_{q,t_i}^R$.  This virtual point keeps static when the Rx is moving. For the initial time and position, we have
\begin{equation}
\textbf{D}_{1,1,t_0n} = \left[ 
\begin{matrix}
D_{1,1,t_0,n}\cos\varphi_{p,q,t_0,n}^E\cos\varphi_{p,q,t_0,n}^A\\
D_{1,1,t_0,n}\cos\varphi_{p,q,t_0,n}^E\sin\varphi_{p,q,t_0,n}^A\\
D_{1,1,t_0,n}\sin\varphi_{p,q,t_0,n}^E
\end{matrix}
\right]. 
\end{equation}
The new distance $D_{t+\Delta t,p,q,f,n}$at $t+\Delta t$ of $\text{Cluster}_n$ can be calculated by 
\begin{equation}
\begin{split}
D_{t+\Delta t,p,q,f,n} &= \left \|\tilde{\textbf{D}}_{p,q,t_i+\Delta t,n}\right \|\\
&=\left \|\tilde{\textbf{D}}_{p,q,t_i,n}+\textbf{v}^R\Delta t\right \|.
\end{split}
\end{equation}

The EAoA and AAoA at $t_i+\Delta t$ $\phi_{p,t,n}^E$ and $\phi_{p,t,n}^A$ equal to elevation and azimuth angle of the virtual vector $\tilde{\textbf{D}}_{p,q,t_i+\Delta t,n}$, respectively. Then, the EAoD and AAoD can be calculated as
 \begin{align}
 \phi_{p,q,t_i+\Delta t,n}^E&=\phi_{p,q,t_i,n}^E+\varphi_{p,q,t_i+\Delta t,n}^E-\varphi_{p,q,t_i,n}^E\\
 \phi_{p,q,t_i+\Delta t,n}^A&=\phi_{p,q,t_i,n}^A+\varphi_{p,q,t_i+\Delta t,n}^A-\varphi_{p,q,t_i,n}^A.
 \end{align}
 
The power of the cluster is also re-calculated according to the updated distance at $t+\Delta t$. Similarly, the phase of the cluster is also updated according to the updated distance. However, the relative delay and relative angle dispersion is considered invariant during time evolution.     

\subsubsection{Evolution in the Space Domain}
The process of evolution in the space domain is similar to the algorithm in the time domain. Evolution in the space domain is also calculated by geometric relationship. The 3D distance between adjacent elements replaces the mobility in the time domain.

%equation

\subsubsection{Evolution in the Frequency Domain}
The scattering from one surface does not keep the same at different frequencies because of different wavelengths. All the intra-cluster parameters including relative delay and relative angle need to be updated when frequency changes. %In this model we assume that variance of relative angle to the cluster center is changing for different frequencies. 
Relative delay is re-generated with new $\mu_{f_i,n}$ by
\begin{equation}
\mu_{f_i,n}=\mu_{f_0,n}\times(\frac{f_i}{f})^{\rho_\mu}.
\end{equation}

Similarly, relative angles need to be generated with the $\sigma_{f_i,n}$ by \begin{equation}
\sigma_{f_i,n}=\sigma_{f_0,n}\times(\frac{f_i}{f})^{\rho_\sigma}.
\end{equation} 
Then, the channel coefficient at frequency $f_i$ can be generated.

\section{Statistical Properties}
In this section, some typical statistical properties of the proposed non-stationary theoretical THz channel model are derived. 

\subsection{The Delay PSD}
The delay PSD $\Upsilon_{p,q,t_i,f_i}(\tau)$ can be written as 
\begin{equation}\label{PDP}
\begin{split}
\Upsilon_{p,q,t_i,f_i}(\tau)&= \left|h_{p,q,t_i}^{\text{LOS}} \right| ^2\times\delta(\tau-\tau_{p,q,t_i}^{\text{LOS}})\\
&+\sum_{n=1}^{N}\sum_{m_n=1}^{M_n}\left| h_{p,q,t_i,f_i,n,m_n}\right|^2\\&\times\delta(\tau-\tau_{p,q,t_i,n}-\tau_{p,q,t_i,f_i,n,m_n}).
\end{split}
\end{equation}
This cluster power at different delays can be calculated from the channel impulse response (CIR) $h_{p,q,t_i,f_i,n,m_n}\delta(\tau-\tau_{p,q,t_i,n}-\tau_{p,q,t_i,f_i,n,m_n})$, which is the inverse Fourier transform of CTF.

%\subsection{Delay Spread}
%Due to the multipath effect of the proposed wideband model, the time of arrival (ToA) is dispersive. The average delay
%$\mu_\tau$ and the root mean square (RMS) delay spread $\sigma_\tau$ arecalculated to describe the delay characteristics of the THzindoor channel. They can be expressed as
%\begin{equation}
%\mu_\tau=\frac{\sum_{\tau}{\tau \phi(\tau)}}{\sum_{\tau}{\phi(\tau)}}
%\end{equation}
%\begin{equation}
%\sigma_\tau=\sqrt{\frac{\sum_{\tau}{(\tau-\mu_\tau)^2 %\phi(\tau)}}{\sum_{\tau}{\phi(\tau)}}}
%\end{equation}

\subsection{Space-Time-Frequency Correlation Function}
%The correlation function of two arbitrary CTF $H_{p_1,q_1,t_1,f_1}(f)$ and $H_{p_2,q_2,t_2,f_2}(f)$ can be defined as the summation of all the
%clusters with no inter-correlation. 
%To investigate the correlation properties, 
The space-time-frequency correlation function
$R_{p,q,t_i,q_i}(\Delta p,\Delta q,\Delta t,\Delta f)$ can be calculated as
\begin{equation}
\label{STF_correlation_function}
\begin{split}
&R_{p,q,t_i,f_i}(\Delta p,\Delta q,\Delta t,\Delta f)\\
= &E\left[H_{p,q,t_i,f_i}(f)\cdot H_{p+\Delta q,q+\Delta p,t_i+\Delta t,f_i+\Delta f}^*(f) \right] \\
= &\frac{K}{K+1}R_{p,q,t_i,f_i}^{\text{LOS}}(\Delta p,\Delta q,\Delta t,\Delta f)\\&+\frac{1}{K+1}R_{p,q,t_i,f_i}^{\text{NLOS}}(\Delta p,\Delta q,\Delta t,\Delta f)
\end{split}
\end{equation}
where $E[\cdot]$ denotes the expectation operator, $(\cdot)^*$ denotes the
complex conjugate operation, $K$ is the Ricean factor which is the ratio of the LOS power to the NLOS power and it can be calculated from the CTF. The correlation function of the channel consists of the LOS and NLOS components. 
%The LoS component is calculated by the locations of the Tx and Rx. The NLoS components are generated randomly. Assuming that the LoS and NLoS components are independent with each other. 

--In the LOS case,
\begin{equation}
\begin{split}
&R_{p,q,t_i,f_i}^{\text{LOS}}(\Delta p,\Delta q,\Delta t,\Delta f)\\
=&E\left[H_{p,q,t_i,f_i}^{\text{LOS}}(f)\cdot H_{p+\Delta q,q+\Delta p,t_i+\Delta t,f_i+\Delta f}^{\text{LOS}^*}(f) \right].
\end{split}
\end{equation}

--In the NLOS case, 
\begin{equation}
\begin{split}
&R_{p,q,t_i,q_i}^{\text{NLOS}}(\Delta p,\Delta q,\Delta t,\Delta f)\\
=&E\bigg[\sum_{n=1}^{N}\sum_{m_n=1}^{M_n}H_{p,q,t_i,f_i,n,m_n}(f) \\&\times H_{p+\Delta q,q+\Delta p,t_i+\Delta t,f_i+\Delta f,n,m_n}^*(f)\bigg].
\end{split}
\end{equation}

By setting partial parameters of ($\Delta p,\Delta q,\Delta t$, $\Delta f$) as 0, the space-time-frequency correlation function can easily be simplified to FCF, time ACF and spatial CCF, which can be expressed as
\begin{align}
R_{p,q,t_i,f_i}^{\text{FCF}}(\Delta f)&= R_{p,q,t_i,f_i}(0,0,0,\Delta f)\\
R_{p,q,t_i,f_i}^{\text{ACF}}(\Delta t)&= R_{p,q,t_i,f_i}(0,0,\Delta t,0)\\
R_{p,q,t_i,f_i}^{\text{CCF}}(\Delta p,\Delta q)&= R_{p,q,t_i,f_i}(\Delta p,\Delta q,0,0).
\end{align}

\subsection{The Stationary Intervals in Space-Time-Frequency Domain}

The stationary interval is the period during which the channel statistical properties can be seen as unchanged. To obtain the stationary interval in space-time-frequency domain, the time-variant correlation matrix distance (CMD) can be applied \cite{RN516}. The CMD can be calculated in space-time-frequency domain as follows
\begin{equation}
\begin{split}
&d_{\text{corr}}(\Delta p_s,\Delta q_s,\Delta t_s,\Delta f_s)\\
&=\frac{tr\left\lbrace R_{p,q,t_i,f_i}R_{p+\Delta p_s,q+\Delta q_s,t_i+\Delta t_s,f_i+\Delta f_s} \right\rbrace }{\left\|R_{p,q,t_i,f_i} \right\|\left\|R_{p+\Delta p_s,q+\Delta q_s,t_i+\Delta t_s,f_i+\Delta f_s} \right\|  }
\end{split}
\end{equation}
where $R_{p,q,t_i,f_i}$ is the correlation function of channel transfer function. The stationary interval can expressed as 
\begin{equation}
\begin{split}
&RG(p,q,t_i,f_i)=\\
&\text{min}\{ \Delta p_s, \Delta q_s, \Delta t_s, \Delta f_s\mid_{d_{\text{corr}}(\Delta p_s,\Delta q_s,\Delta t_s,\Delta f_s) \geqslant c_{th}} \}
\end{split}
\end{equation}
where $\Delta p_s$ and $\Delta q_s$ are the space stationary interval at Tx and Rx,respectively. $\Delta t_s$ is the time stationary interval, and  $\Delta f_s$ is the frequency stationary. $c_{th}$ is the threshold which can be adjusted in different cases. The above stationary intervals can be used to evaluate the non-stationary behaviors of THz channels in space/time/frequency domains. 
%\subsection{The Stationary Intervals in Space-Time-Frequency Domain}

%The stationary interval is the period during which the channel can be seen as unchanged compared with the neighbor channel. To obtain the stationary interval in space-time-frequency domain, the time-variant correlation matrix distance (CMD) can be applied \cite{RN516}. The CMD can be calculated in space-time-frequency domain as follows
%\begin{equation}
%\begin{split}
%&d_{\text{corr}}(\Delta p_s,\Delta q_s,\Delta t_s,\Delta f_s)\\
%&=1-\frac{tr\left\lbrace R_{p,q,t_i,f_i}R_{p+\Delta p_s,q+\Delta q_s,t_i+\Delta t_s,f_i+\Delta f_s} \right\rbrace }{\left\|R_{p,q,t_i,f_i} \right\|\left\|R_{p+\Delta p_s,q+\Delta q_s,t_i+\Delta t_s,f_i+\Delta f_s} \right\|  }
%\end{split}
%\end{equation}
%where $R_{p,q,t_i,f_i}$ is the correlation function of channel transfer function. $\Delta p_s$ and $\Delta q_s$ are the space stationary interval at Tx and Rx,respectively. $\Delta t_s$ is the time stationary interval, and  $\Delta f_s$ is the frequency stationary. The above stationary intervals can be used to evaluate the non-stationary behaviors of THz channels in space/time/frequency domains. 

\section{Results and Discussions}
In this part, the statistical properties of the proposed THz channel models are studied and analyzed. The antenna arrays of Tx and Rx are assumed to be uniform linear array (ULA). The element of the array is assumed as omnidirectional and the gain is 1 in all directions. The related parameters are listed as follows. The moving speed of Rx is $\text{v}^R$ = 0.1 m/s with the direction angle $\theta^E$ = 0 and $\theta^A$~=~$\frac{\pi}{3}$ while the Tx is fixed. The frequency band is chosen from 300 GHz to 400 GHz. The numbers of antenna elements at the Tx and Rx were both set as 1024. The initial distance between the first elements of Tx and Rx is 3 m, $\mu_{\Delta \tau_{i,1st}}$ and $\mu_{\Delta \tau_{i,2nd}}$ are set as 2.73 ns and 4.8 ns, respectively. All the initial angle parameters are generated by Gaussian
distribution. The standard deviations of $\phi_{1,1,t_0,n}^A$, $\phi_{1,1,t_0,n}^E$,$\varphi_{1,1,t_0,n}^A$, and $\varphi_{1,1,t_0,n}^E$ are set as 1.2. The $\rho_\mu$ and $\rho_\sigma$ are set as 3. The number of rays in each cluster is set as 100.
% The statistical properties of simulation model are  obtained from ten thousand times of Monte Carlo simulations.

\subsection{ACF}
\begin{figure}[tb]
	\centerline{\includegraphics[width=0.47\textwidth]{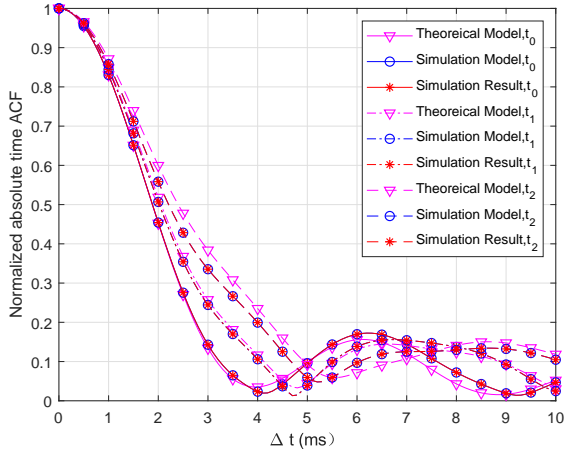}}
	\caption{The comparison of time-variant ACF of theoretical model, simulation model, and simulation results at $t_0$=0 s, $t_1$=5 s, and $t_2$=10 s (p=1, q=1, $f_i$=300 GHz, $\text{v}^R$=0.1 m/s).}
	\label{fig_ACF}
\end{figure}

By setting $\Delta p,\Delta q,\Delta f$ as 0, the time ACF of the theoretical model can be obtained. 
%After ten thousand times of Monte Carlo simulations, statistical proporties of 
The comparison of theoretical model, simulation model, and simulation result at  $t_0$=0 s, $t_1$=5~s, and $t_2$=10~s of $\text{Cluster}_1$ is shown in Fig. \ref{fig_ACF}. From the results, we can see that the simulation model provides a good approximation to the theoretical model. We can observe different time ACFs at different time instants, which demonstrates that the proposed model can capture the non-stationarity of channel in the time domain.

\subsection{Spatial CCF}
\begin{figure}[tb]
	\centerline{\includegraphics[width=0.47\textwidth]{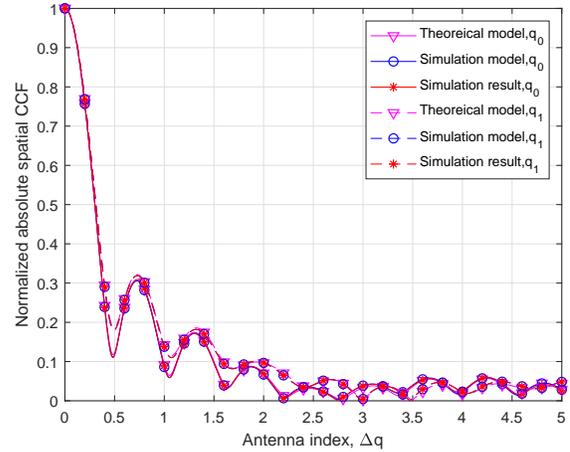}}
	\caption{The comparison of spatial CCFs of theoretical model, simulation model, and simulation results at $q_0$=1,$q_1$=1000 (p=1, $t_0$=0s, $f_i$=300 GHz, $\text{v}^R$=0.1 m/s).}
	\label{fig_CCF}
\end{figure}
The comparison of the theoretical model, simulation model, and simulation result for $\text{Cluster}_1$ with different receive elements is shown in Fig. \ref{fig_CCF}. we can see that the differences of two spatial CCFs are quite similar because the antenna spacing is not large because the antenna spacing is quite small in THz band. However, the differences are still observable. If the antenna size is large enough, non-stationarity of THz channel in the space domain should be considered. 
\begin{figure}[t]
	\centerline{\includegraphics[width=0.47\textwidth]{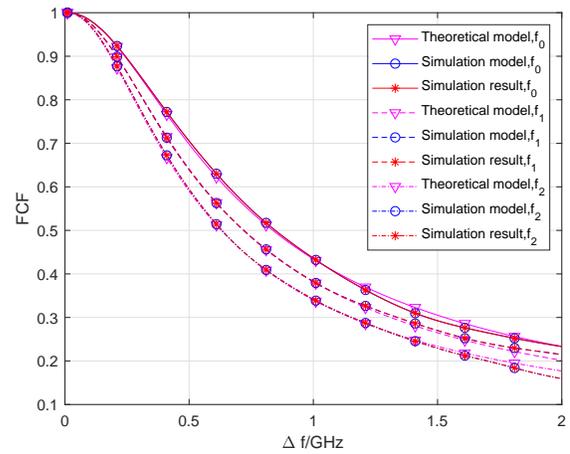}}
	\caption{The comparison of FCFs of theoretical model, simulation model, and simulation results at $f_0$=300 GHz, $f_1$=325 GHz, and $f_2$=350 GHz (p=1, q=1, $t_0$=0 s, $\text{v}^R$=0.1 m/s).}
	\label{fig_FCF}
\end{figure}

\subsection{FCF}
The results of FCF at different  $f_0$=300 GHz, $f_1$=325 GHz, and $f_2=350$ GHz are shown in Fig.~\ref{fig_FCF}. We can observe the differences of FCF at different frequencies are small clearly in the picture. Due to the frequency dependent intra-cluster parameters, the channel can not be considered as stationary in frequency domain. 
%If the communication frequency is up to 1THz, the cha

\begin{figure}[t]
	\centerline{\includegraphics[width=0.5\textwidth]{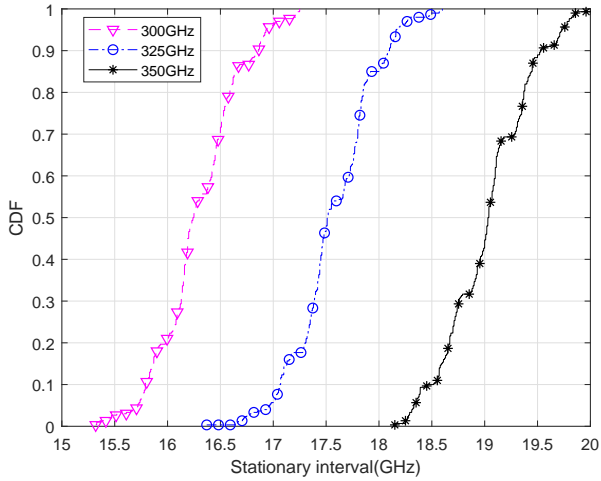}}
	\caption{The comparison of stationary interval of the simulation model in frequency domain at $f_0$=300 GHz, $f_1$=325 GHz and $f_2$=350 GHz (p=1, q=1, $t_0$=0 s, $\text{v}^R$=0.1 m/s), $c_{th}=0.9$.}
	\label{Stationinterval}
\end{figure}
\subsection{Stationary Interval}
The stationary intervals of the simulation model are obtained by 5000 times of Monte Carlo simulations. The CCDFs of the stationary intervals in frequency domain of the proposed channel model in different frequencies are shown in Fig. \ref{Stationinterval}. The threshold is set as 0.9.  The channel whose bandwidth less than the stationary interval can be considered as frequency stationary. For higher frequency, the frequency stationary interval will linearly increase.

\section{Conclusions}
In this paper, a novel 3D space-time-frequency non-stationary THz massive MIMO channel model for 6G THz indoor communication systems has been proposed. The initialization and evolution of parameters in time, space, and frequency domains have been given to generate the complete channel. Based on the proposed  models, the %cluster level
correlation functions including ACF, CCF, and FCF have been investigated. Numerical and simulation results have shown that the statistical properties of the simulation model match well with those of the theoretical model. The non-stationarity in time, space, and frequency domains have been verified by theoretical derivations and simulations.

%Moreover, due to the movements of clusters and MRS, the statistical properties experience different behaviors at different time instants, which has demonstrated that our proposed models can mimic the non-stationarity of HST channels. By introducing the model parameters in space, time, and frequency domains, the joint space-time-frequency non-stationarity of the proposed HST channel model has been investigated.

\section*{Acknowledgment}
\small {This work was supported by the National Key R\&D Program of China under Grant 2018YFB1801101, the National Natural Science Foundation of China (NSFC) under Grant 61960206006 and 61901109, the Research Fund of National Mobile Communications Research Laboratory, Southeast University, under Grant 2020B01, the Fundamental Research Funds for the Central Universities under Grant 2242019R30001, National Postdoctoral Program for Innovative Talents (No. BX20180062), and the EU H2020 RISE TESTBED2 project under Grant 872172.}

\bibliographystyle{IEEEtran}
\bibliography{WCNC20_THz_channel.bib}

\end{document}